\documentclass{article} 

\usepackage{amsmath,amssymb,amsthm,latexsym} 

\usepackage{stmaryrd,wasysym,upgreek,mathrsfs,dsfont} 
\usepackage[english]{babel} 
\usepackage{graphicx,color} 
\newtheorem{lemma}{Lemma}[section]

\newcommand{\bbbone}{{\mathds{1}}}
\newtheorem{definition}{Definition}[section]
\newtheorem{theorem}{Theorem}[section]
\newcommand{\be}{\begin{equation}}
\newcommand{\ee}{\end{equation}}
\newcommand{\bea}{\begin{eqnarray}}
\newcommand{\eea}{\end{eqnarray}}

\newcommand{\Tr}{{\rm Tr}}
\newcommand{\cB}{{\cal{B}}}
\newcommand{\cD}{{\cal{D}}}

\newcommand{\cL}{{\cal{L}}}

\newcommand{\cT}{{\cal{T}}}
\newcommand{\cP}{{\cal{P}}}
\newcommand{\cF}{{\cal{F}}}

\begin{document} 
%\begin{titlepage}

 \title{Loop Vertex Expansion\\
  for $\Phi^{2k}$ Theory in Zero Dimension}
\author{Vincent Rivasseau,\  Zhituo Wang\\
Laboratoire de Physique Th\'eorique, CNRS UMR 8627,\\ 
Universit\'e Paris XI,  F-91405 Orsay Cedex, France}

\maketitle 
\begin{abstract} 
In this paper we extend the method of loop vertex expansion 
to interactions with degree higher than 4. As an example we provide
through this expansion an explicit proof that the free energy of  $\phi^{2k}$ scalar theory in zero dimension is Borel-Le Roy summable of order $k-1$. We detail the computations in the case of 
a $\phi^{6}$ interaction.

\end{abstract} 

\begin{flushright}
LPT-20XX-xx
\end{flushright}
\medskip


\medskip

\section{Introduction} 

New constructive Bosonic field theory methods have been recently proposed \cite{R1,MR1,GMR}. The
method called Loop vertex expansion or Cactus expansion \cite{R1,MR1,MNRS}.
is based on applying a canonical forest formula to repackage perturbation theory in a better way.
This allows to compute the connected quantities of the theory by the same formula but summed over trees rather than forests.  Combining the forest formula with the intermediate field method leads to a convenient resummation of $\phi^4$ perturbation theory. 

The main advantage of this formalism over previous cluster and Mayer expansions
is that connected functions are captured by a single formula, and e.g. a Borel summability theorem 
for matrix $\phi^4$ models can be obtained which scales correctly with the size of the matrix.

In this paper we extend this method, which at first sight looks limited to $\phi^4$ interactions,
to show that it is in fact suitable for any stable quantum field theory.
For simplicity we restrict ourselves  to interactions of the $\lambda \phi^{2k}$ type
in zero dimension. We introduce several intermediate fields instead of one for the $\phi^4$ model. 
We also take care of the integration contours to bound the integral over intermediate fields.
We prove the Borel-Le Roy summability of the right order for this class of theories.
Extension to quantum field theories in more than 0 dimension in the line of \cite{MR1} is
devoted to a future publication, but should follow from the method of this paper and the 
local nature of the interaction.

%We recommend the reading of \cite{R1,MR1} before going further down this paper.

\section{The Forest Formula}

This formula, a key tool in constructive theory, was perfected along the years by many authors
 \cite{BK,AR1}. It is shown here as a Taylor-Lagrange
expansion, in which a function of many link variables is expanded around the origin in a
careful and symmetric way which stops with an integral remainder 
before the derivatives create any cycles. 

Consider $n$ points. The set of pairs $P_n$ of such points has
$n(n-1)/2$ elements $\ell = (i,j)$ for $1\le i < j \le n$. Consider a smooth function $f$
of $n(n-1)/2$ variables $x_\ell$, $\ell \in \cP_n$. Noting $\partial_\ell$ 
for $\frac{\partial}{\partial x_\ell}$, the forest formula is 
\begin{theorem}

\be\label{treeformul1}
f(1,\dots ,1)
= \sum_{\cF}  \big[ \prod_{\ell\in \cF}   
\int_0^1 dw_\ell   \big] \big( [ \prod_{\ell\in \cF} \partial_\ell ] f 
\big) \cdot [ X^\cF (\{ w_{\ell'}\} ) ]
\ee
where 
\begin{itemize}

\item the sum over $\cF$ is over forests over the $n$ vertices, including the empty one,

\item $x^\cF_\ell (\{ w_{\ell'}\} )$ is the infimum of the $w_{\ell'}$ for $\ell'$
in the unique path from $i$ to $j$ in $\cF$, where $\ell = (i,j)$. If there is no such path,
$x^\cF_\ell (\{ w_{\ell'}\} ) = 0$ by definition.

\item The symmetric $n$ by $n$ matrix $X^\cF (\{w\})$ defined
by $X^\cF_{ii} = 1$ and $X^\cF_{ij} =x^\cF_{ij} (\{ w_{\ell'}\} ) $ 
for $1\le i < j \le n$ is positive.

\end{itemize}
\end{theorem}

\begin{proof}
We do not reproduce here the many proofs of formula (\ref{treeformul1}) \cite{BK}\cite{AR1}, but
we recall the reason for which the matrix $X^\cF (\{w\})$ is positive. It is because for any 
ordering of the $\{ w\}$ parameters it can be written 
as a (different!) convex combination of positive block matrices of the $I_q$ type.
\begin{definition}
A block $I_q$ of dimension $q$  is defined as a $q\times q$ matrix 
with all the elements $1$. For example, a block of dimension $3$ is:
\begin{equation}
 I_3=\begin{pmatrix}
1 & 1 & 1\\
1 & 1 & 1 \\
1 & 1 & 1 \\
\end{pmatrix} .
\end{equation}
\end{definition}

Consider indeed a forest $\cF$ with $p\le n-1$ elements and an ordering
\begin{equation}
0= w_{p+1} \leq w_{p}\leq w_{p-1} \leq...\leq w_{1} \leq w_{0}=1,
\end{equation}
then
\begin{equation}\label{convexkey}
X^\cF(\{w\}) =\sum_{k=1}^{p+1}(w_{k-1}-w_{k})X^{\cF,k} 
\end{equation}
where $X^{\cF,k}_{ij} $ is 1 if $i$ and $j$ are connected by the $k-1$
first lines of the forest, and is 0 otherwise. 
We have \begin{equation}
 \sum_{k=1}^{p+1}(w_{k-1}-w_{k}) = 1 .
\end{equation}

Therefore $X^{\cF,k} $ is a matrix obtained
by gluing the blocks corresponding to the connected components of the forest $\cF^k$, where
$\cF^k$ is the subforest of $\cF$ made of the $k-1$
first lines of the forest in the ordering.
\end{proof}

We need later the fact that the Gaussian measure $d\mu_{I_q} (a_1, ... a_q)$ with covariance $I_q$ really corresponds to a single Gaussian variable, say $a_1$, with covariance 1, plus $q-1$ delta functions:
\be\label{blockvari} d\mu_{I_q} (a_1 ... a_q) =
 \frac{d a_1}{\sqrt{2 \pi}} e^{- a_1^2/ 2} 
\prod_{i=2}^q  \delta(a_{1} - a_i ) da_{i} \; . 
\ee

\section{$\phi^6$ constructive theory in zero dimension}
We consider a massless $\phi^6$ scalar  theory in zero dimension, where $\phi$ is simply a number. The Lagrangian reads:
\begin{equation}
\cL=-\frac{1}{2}\phi^2-\lambda \phi^6
\end{equation}
and the partition function is
\begin{equation}
Z(\lambda)= \int \frac{d\phi}{\sqrt{2\pi}} e^{-\frac{1}{2}\phi^2} e^{-\lambda\phi^6} .
\end{equation}

The  covariance of the normalized Gaussian measure 
$\frac{d\phi}{\sqrt{2\pi}} e^{-\frac{1}{2}\phi^2}$
is simply
\begin{equation}
< \phi^2> =1 .
\end{equation}

\subsection{Intermediate Field Representation}
We introduce a real intermediate field $\sigma$ to rewrite the interaction. This leads to
\begin{equation}
Z(\lambda)= \int\frac{d\phi}{\sqrt{2\pi}} e^{-\frac{1}{2}\phi^2} e^{-\lambda\phi^6}=\int \frac{d\phi}{\sqrt{2\pi}} e^{-\frac{1}{2}\phi^2} \int  \frac{d\sigma}{\sqrt{2\pi}}  e^{-\frac{1}{2}\sigma^2} e^{i\sqrt{2\lambda}\phi^3\sigma}.
\end{equation}
The induced interaction term could be further transformed as
\begin{equation}
 \sqrt{2}\phi^3\sigma= \frac{1}{\sqrt 2} [(\phi\sigma+\phi^2)^2-\phi^2\sigma^2-\phi^4] .
\end{equation}
We then introduce another three intermediate  fields to write the partition function as
\begin{eqnarray}
Z(\lambda) &=&\int \frac{d\phi}{\sqrt{2\pi}} e^{-\frac{1}{2}\phi^2} \int  \frac{d\sigma}{\sqrt{2\pi}}  e^{-\frac{1}{2}\sigma^2} 
\int \frac{da \sqrt{i}}{\sqrt{2\pi}} e^{i[(2\lambda)^{1/4}(\phi\sigma+\phi^2)a-a^2/2]}\nonumber\\
&\times& \int  \frac{db}{\sqrt{2i\pi}}e^{-i[(2\lambda)^{1/4}\phi\sigma b-b^2/2]}\int  \frac{dc}{\sqrt{2i\pi}} 
e^{-i[(2\lambda)^{1/4}\phi^2 c-c^2/2]} .
\end{eqnarray}

Integrating out the fields $\phi$ and $\sigma$ we get:
\begin{equation} \label{imagauss}
Z(\lambda) = \int  \frac{da \sqrt{i}}{\sqrt{2\pi}}   \frac{db}{\sqrt{2i\pi}}  \frac{dc}{\sqrt{2i\pi}}  
e^{i(b^2+c^2-a^2)/2}e^{V}
\end{equation}
where 
\begin{equation}
 V=-\frac{1}{2}\Tr \ln[\bbbone+i(2\lambda)^{1/4}\begin{pmatrix}
c-a &  b-a  \\
b-a & 0\\
\end{pmatrix}]=-\frac{1}{2}\Tr\ln(\bbbone+iH) ,
\end{equation}
where 
\be \bbbone=\begin{pmatrix}
1 & 0\\
0 & 1 \\
\end{pmatrix} , \quad H=(2\lambda)^{1/4}\ \begin{pmatrix}
c-a &  b-a  \\
b-a & 0\\
\end{pmatrix} .
\ee
Obviously $H$ defined above is Hermitian for $\lambda \ge 0$.

The new resulting integrals (\ref{imagauss}) over $a$, $b$ and $c$ are oscillating and still formal, 
and we have to slightly change the contours of integration to make them well-defined, 
but this is postponed to the next section.

We use the replica method to write the exponential as:
\begin{equation}
e^{V}=\sum_{n}\frac{V^n}{n!}=\sum_{n}\frac{1}{n!}\prod_{v=1}^n V_v
\end{equation}
where \begin{equation}
 V_v=V_v (a^v,b^v,c^v) .
\end{equation}

Then applying the forest formula,  the connected function could be written as
a sum over trees $\cT$ whose nodes are loop vertices, and whose lines are
of three different types, corresponding to Wick contractions of $a$, $b$ and $c$.
Calling $\cT_a$, $\cT_b$ and $\cT_c$ the three corresponding subset of 
lines of the tree we have
\begin{theorem}
\begin{eqnarray}\label{treeformul}
\log Z(\lambda) &=& 
\sum_{n=1}^{\infty}\frac{1}{n!}\ \sum_{\cT \; {\rm with }\; n\; {\rm vertices}}\ Y_\cT
\\
Y_\cT &=&
\bigg\{ \prod_{\ell\in \cT}   
\big[ \int_0^1 dw_\ell  \big]\bigg\} 
\int  d\nu_\cT (\{a^v,b^v,c^v\}, \{ w \})  \nonumber \\
&\times& \bigg\{ \prod_{\ell\in \cT_a} \big[ \delta _{v,v'}
 \frac{\delta}{\delta a^{v(\ell)}}
 \frac{\delta}{\delta a^{v'(\ell)}} 
\big] \bigg\}  \bigg\{ \prod_{\ell\in \cT_b} \big[ \delta _{v,v'} 
 \frac{\delta}{\delta b^{v(\ell)}}
 \frac{\delta}{\delta b^{v'(\ell)}} 
\big] \bigg\}\nonumber\\
&\times& \bigg\{ \prod_{\ell\in \cT_c} \big[ \delta _{v,v'}
 \frac{\delta}{\delta c^{v(\ell)}}
 \frac{\delta}{\delta c^{v'(\ell)}} 
\big] \bigg\}\prod_{v=1}^n V_v 
\end{eqnarray}
where
\begin{itemize}

\item each line $\ell$ of the tree joins two different loop
vertices $V^{v(\ell)}$ and $V^{v'(\ell)}$,

\item the sum is over trees joining $n$ loop vertices, which have therefore
$n-1$ lines. These lines can be of type $a$, $b$ or $c$.

\item the normalized ``imaginary'' Gaussian 
measure $d\nu_T (\{a^v, b^v, c^v\}, \{ w \})  $ over the three intermediate fields $a^v$, $b^v$ and $c^v$ has covariance 
\begin{eqnarray}<a^v, a^{v'}>&=& - i w^T (v, v', \{ w\}),\\
<b^v, b^{v'}>&=&  i w^T (v, v', \{ w\}),\\
<c^v, c^{v'}>&=&  i w^T (v, v', \{ w\}),\\
<a^v, b^v>&=&<b^v, c^v>\; = \; <a^v, c^v>\;=\; 0 
\end{eqnarray}
where $w^T (v, v', \{ w\})$ is 1 if $v=v'$,
and the infimum of the $w_\ell$ for $\ell$ running over the unique path from $v$ to $v'$ in $T$
if $v\ne v'$. This measure will become well-defined since the matrix $w^T$ is positive,
if we perform appropriate contour deformations.

\end{itemize}
\end{theorem}

If we distinguish the matrix indices which correspond to the former 
$\phi$ and $\sigma$ fields, there are in fact four kinds of half-vertices in the 
loop vertex expansion and five different kinds of lines.
The coupling constant for each half-vertex is $(2\lambda)^{1/4}$,
and the coupling constant for each vertex (namely each line of the loop vertex tree) is therefore $(2\lambda)^{1/2}$.
\begin{figure}[!htb]\label{halfv}
\centering
\includegraphics[scale=0.8]{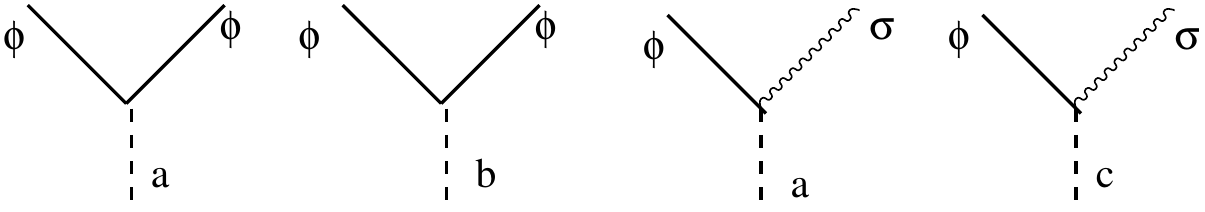}
\caption{The 4 half-vertices}
\label{vertex}
\end{figure} 
\begin{figure}[!htb]\label{fulver}
\centering
\includegraphics[scale=0.8]{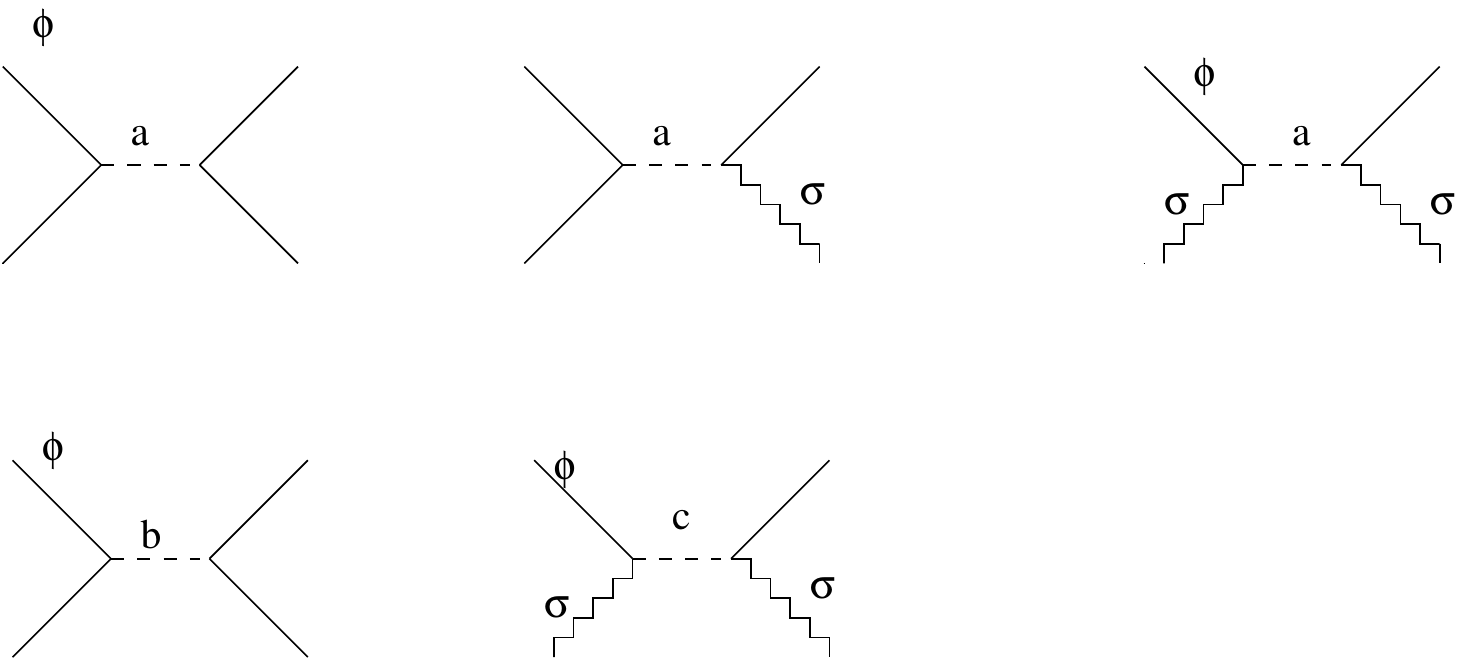}
\caption{The 5 vertices}
\label{vertex5}
\end{figure}

\subsection{Contour Deformation} \label{contourdefo}
The integral over the fields $a$, $b$ and $c$ is not absolutely 
convergent, so we have to choose the right contour to make it 
well-defined. As the covariance for the three fields are quite similar, we will consider $a$ first
and deform the integration contour. The idea is that, we first of all use the formula (\ref{convexkey})
to write the field $a$ as an independent sum of $p+1$ fields $a_{k}$ according to the blocks:
\begin{equation}
 a=\sum_{k=1}^{p+1}\sum_{v=1}^n a_{k,v}
\end{equation}
whose covariance is
\be  <a_{k,v},  a_{k,v'}>  =  (w_{k-1} - w_k)  X^{\cF, k}_{v,v'}  .
\ee
Precisely because the covariance of $a_k$ is made of blocks $r$, we should perform a single
contour deformation for each block.
We have a formula similar to (\ref{blockvari}) for each  
block with variables $a_1, ... a_n$, but now we should remember that the covariances are 
$iI_q$, not $I_q$. Hence we have
\be \label{singlevar} d\mu_{iI_q} =
 \frac{d a_1}{\sqrt{2 \pi}} e^{- i a_1^2/ 2} 
\prod_{i=2}^q  \delta (a_{1} - a_i ) da_{i} .
\ee

In the partition function we have an integration of the type
\begin{equation}
 \int_{-\infty}^{\infty} da f(a) e^{ia^2 /2}
\end{equation}
where $f$ is the product of the resolvents which are analytic and 
bounded in a open neighborhood of the band $\cB= \{ \Im a \le A^{-1} \}$ 
of the real axis, where $A$ is large.
This integral is not absolutely convergent. Nevertheless we can bound it 
in terms of $\sup_\cB \vert f\vert$. Indeed we can deform the integral contour
so that the new contour remains in the band $\cB$ and the new variable is:
\begin{eqnarray}
 a'_1&=&a_1-i\frac{a_1}{A|a_1|+1}, \ a'_1 \to a_1 - i {\rm sgn}\; a_1/A \   {\rm if}\  a_1\to \pm\infty  .%\nonumber \\
\end{eqnarray}
Then the bound of the integral over $a_1$ becomes:
\bea
&&\bigg|  \int d a_1 f(a'_1)  e^{-ia_1^2/2(w_{k-1}-w_k)-\frac{2 a_1^2}{2(w_{k-1}-w_k)(A|a_1|+1)}  +i  \frac{a_1^2}{2(w_{k-1}-w_k)(A|a_1|+1)^2}}\bigg| \nonumber
\\  \nonumber
&& \le   \sup_\cB\vert f \vert \int d a_1 e^{-\frac{2 a_1^2}{2(w_{k-1}-w_k)(A|a_1|+1)} }
\\
&& \le 
2 (w_{k-1}-w_k)A\;  \sup_\cB\vert f \vert  . \label{mainwbound}
\eea
So each time we integrate out an intermediate field we get $\sup_\cB \vert f \vert$ 
times a factor $2(w_{k-1}-w_k)A$ 
in the bound. Then for the integration of all the intermediate fields $a_k$ 
we would have at order $n$ a total factor in the bound:
\begin{eqnarray}
 &&\prod_{k=1}^{p+1}\prod_{v=1}^n 2A(w_{k-1}-w_k)\le\prod_{k=1}^{p+1} \prod_{v=1}^n e^{ 2A(w_{k-1}-w_k)}=\prod_{v=1}^n e^{\sum_{k=1}^{p+1} 2A(w_{k-1}-w_k)}\nonumber\\
&\le&\prod_{v=1}^n e^{2A}\le (e^{2A})^n  \label{contourbound}
\end{eqnarray}
 where we have used the fact that 
\begin{equation}
\sum_k (w_{k-1}-w_k)\le 1 .
\end{equation}

\begin{figure}[!htb]
\centering
\includegraphics[scale=0.8]{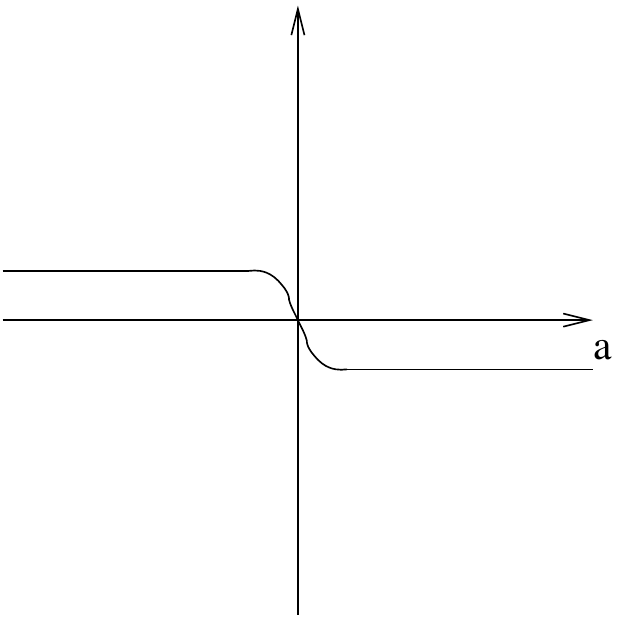}
\caption{The integral contour for a.}
\label{contour}
\end{figure}
As the signs for $b$ and $c$ are different from $a$ in the covariance, the integral contour for $b$ and $c$ are also different. The contour for $b$ could be chosen as:
\begin{eqnarray}
 b'_1&=&b_1+i\frac{b_1}{A|b_1|+1}, \  b'_1 \to b_1 + i {\rm sgn}\; b_1/A \  
 {\rm  if}\  b_1\to \pm\infty  \nonumber \\
\end{eqnarray}
and the integral contour for $c$ is the same as that for $b$. Then the bounds proceed exactly like
in (\ref{mainwbound}).
\begin{figure}[!htb]
\centering
\includegraphics[scale=0.8]{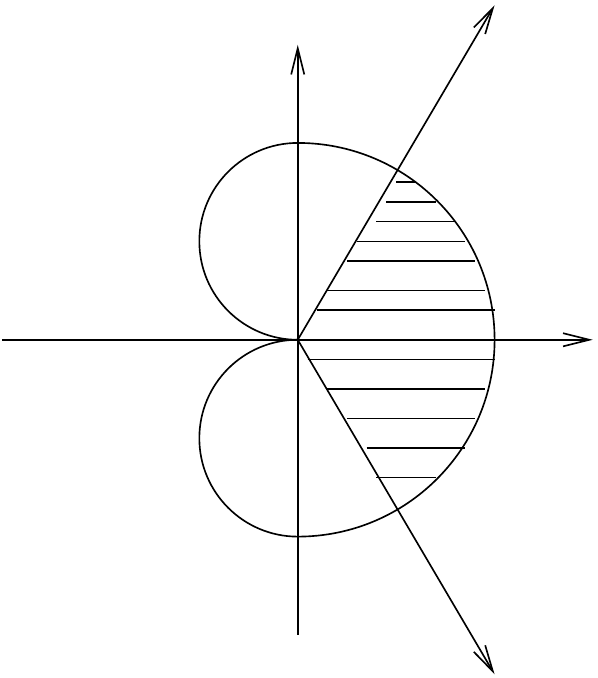}
\caption{The analyticity domain $C^2_R$}
\label{borel1}
\end{figure}
\begin{figure}[!htb]
\centering
\includegraphics[scale=0.8]{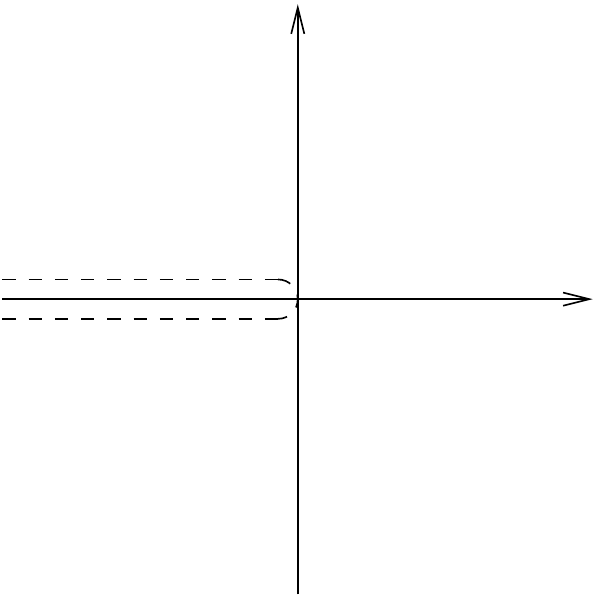}
\caption{The analyticity domain $\cD^2$}
\label{borel2}
\end{figure}
\begin{figure}[!htb]
\centering
\includegraphics[scale=0.8]{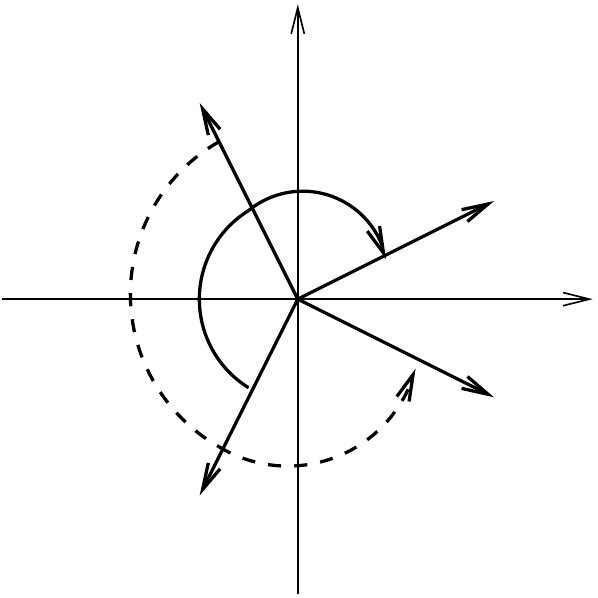}
\caption{The analytic continuation}
\label{borel3}
\end{figure}

The function $f$ is a product of resolvents of the type $(1+iH)^{-1}$ turning 
around the tree after using the tree formula \cite{R1}.  On the real axis
$\Vert (1+iH)^{-1}\Vert \le 1$. But after contour deformation the bound is slightly altered.
$1+iH$ will also be changed into 
\begin{equation}
 1+iH-(2\lambda)^{1/4}\begin{pmatrix}
\epsilon(a) &  \epsilon(a)  \\
\epsilon(a) & 0\\
\end{pmatrix}
\end{equation}
with $\epsilon = 1/A$ a small number.
As $\lambda<<1$, $\Vert {\lambda}^{1/4}\begin{pmatrix}
\epsilon(a) &  \epsilon(a)  \\
\epsilon(a) & 0\\
\end{pmatrix}\Vert <<1$.  So after we change the integral contours,
the denominators are still bounded by $K= 1 + O(1/A)$. This bound changes
to $\sqrt2 + O(1/A)$ if we take $-\pi < Arg \lambda <+\pi$, that is $-\pi/4 < Arg \lambda^{1/4} <+\pi/4$.
As essentially the factor $O(1/A)$ doesn't change the bound of the resolvent, hence the power counting of the connected function, we shall forget it in the rest of this paper.

As $H$ is a linear function of $b$ and $c$, we could use the same method for $b$ and $c$ and  the resulting integral is finite. 

\subsection{Borel summability}

Let us introduce the $N$-th order Taylor remainder operator $R^N$ which acts on a function $f(\lambda)$ through
\bea  R^N  f = f(\lambda)-\sum_{n=0}^{N} a_n \lambda^n  = \lambda^{N+1}
\int_0^1 \frac{(1-t)^{N}}{N!} f^{(N+1)} ( t \lambda) dt .
\eea

\begin{theorem}(Nevanlinna-Le Roy)\cite{Borel, Sok}

A series $\sum_{n=0}\frac{a_n}{n!}\lambda^n$ is Borel summable to the function $f(\lambda)$ of order $k$ if the following conditions are met:

\begin{itemize}
 \item For some rational number $k>0$, $f(\lambda)$ is analytic in the domain $C^k_R=\{\lambda\in C: 
\Re \lambda^{-1/k}> R^{-1}\}$. $C_R$ is a disk for $k=1$.
\item The function $f(\lambda)$ admits $\sum_{n=0}^\infty a_n \lambda^n$ as a strong asymptotic expansion to all orders as $|\lambda|$ $\rightarrow 0$ in $C_R$ with uniform estimate in $C^k_R$:
\begin{equation}
\left|  R^N f  \right|\leqslant A B^N \Gamma(kN+1)|\lambda|^{N+1}.
\end{equation}
where $A$ and $B$ are some constants.
\end{itemize}
Then the Borel-Le Roy transform of order $k$ reads:
\begin{equation}
B^{(k)}_f(u)=\sum_{n=0}^\infty \frac{a_n}{\Gamma(kn+1)}u^n,
\end{equation}
it is holomorphic for $|u|<B^{-1}$, it admits an analytic continuation to the strip 
$\{u\in C: |\Im u|< R, \Re u>0\}$ and for $0\leqslant R$, one has
\begin{equation}
 f(\lambda)=\frac{1}{k\lambda}\int_{0}^{\infty}B^{(k)}_f (u) exp[-(u/\lambda)^{1/k}](u/\lambda)^{(1/{k-1})}du .
\end{equation}

\end{theorem}

\begin{theorem}
The partition function $Z(\lambda)$ for $\phi^6$ theory is Borel-Le Roy summable
of order 2.\label{th1}
\end{theorem}

\begin{proof}
The remainder after the Taylor expansion of $\phi^6$ at $N$th order reads:
\begin{equation}
R^N  Z(\lambda)=(-\lambda)^{N+1} 
\int_0^1dt\int d\phi\frac{(1-t)^N}{N!}\phi^{6(N+1)}e^{-t\lambda\phi^6-\frac{\phi^2}{2}} .
\end{equation}

We use the Cauchy-Schwarz inequality:
\begin{eqnarray}
&& |R^{N}  Z(\lambda)|= \vert \lambda \vert^{N+1} \int_0^1dt\int d\phi\frac{(1-t)^N}{N!}[\phi^{12(N+1)}e^{-2t\lambda\phi^6-{\phi^2}}]^{1/2}
\\&\leqslant&\vert \lambda \vert^{N+1}  \int_0^1dt\frac{(1-t)^N}{N!}(\int d\phi \phi^{12(N+1)}
e^{-\phi^2/2})^{1/2}
(\int d\phi e^{-2t\lambda \phi^6}e^{-\phi^2/2})^{1/2} .\nonumber
\end{eqnarray}
The first term is bounded by $[(12(N+1))!!]^{1/2}/N!\sim {(6N)!!}/N!\sim (2N)!$, where $\sim \cdots$ means $\le K^N \times \cdots$.
Now consider the second term. We perform a scaling on $\phi$ as 
\begin{equation}
 \lambda^{1/6}\phi=u
\end{equation}
then
\begin{equation}\label{analytrem}
\int d\phi e^{-2t\lambda \phi^6}e^{-\phi^2/2}=\int_{-\infty}^{\infty}e^{-2t u^6-\lambda^{-1/3}u^2}\frac{du}{\lambda^{1/6}} .
\end{equation}
For $-\pi < Arg{\lambda}< \pi$,  we have  $-\pi/3 < Arg(\lambda^{1/3})< \pi/3$. 
Let us define $\cD^2 = \{ \lambda \vert -\pi < Arg{\lambda}< \pi  \}$. We have 
$C^2_R \subset \cD^2 $.
The corresponding analytic domains are shown in figure \ref{borel1} and \ref{borel2}. 
We shall prove analyticity and Taylor remainder bounds in $\cD^2$ rather than $C^2_R$.

In $\cD^2$  the integrand of (\ref{analytrem}) is analytic in $\lambda$ and we always have $\Re \lambda^{-1/3}>0$. Moreover
we have uniform convergence
\begin{equation}
\vert \int_{-\infty}^{\infty}e^{-2t u^6-\lambda^{-1/3}u^2}\frac{du}{\lambda^{1/6}}\vert \le 
 \int_{-\infty}^{\infty}e^{- (\Re \lambda^{-1/3} ) u^2}\frac{du}{ \vert \lambda^{1/6} \vert } \le \sqrt \pi .
\end{equation}

This proves that the partition function is Borel summable of order $2$.
\end{proof}

The rest of this section is devoted to prove the following more difficult results:
\begin{theorem}\label{theoborel3}
The connected function $\log Z(\lambda)$ with potential $\lambda\phi^{6}$ 
is Borel-Le Roy summable of order $2$.
\end{theorem}
\begin{proof}
We use the loop vertex representation (\ref{treeformul}) of $\log Z(\lambda)$.
We shall first prove uniform convergence of this loop vertex representation
in the domain $\cD^2_{\epsilon}=  \{ \lambda \vert -\pi < Arg{\lambda}< \pi 
\ {\rm and} \ \vert \lambda \vert < \epsilon \}$ and then prove the Taylor remainder bound.

\begin{lemma}
In the domain $\cD^2_{\epsilon}$ each term $Y_\cT (\lambda)$ is bounded by 
$ \epsilon^{(n-1)/2} K^n$\label{lem1}.
\end{lemma}
\begin{proof}
In the loop vertex expansion remember there are 4 different kinds of half- vertices, 
as shown in Figure \ref{halfv}, and five different types of tree lines after contraction of the $a$ $b$ or $c$ intermediate fields, as shown in Figure (\ref{fulver}).

We shall first of all prove that the resolvents are bounded. Consider
 \begin{equation}
\frac{1}{1+i H}=\frac{1}{\begin{pmatrix}
1 &  0  \\
0 & 1\\
\end{pmatrix}+i(2\lambda)^{1/4}
\begin{pmatrix}
c-a &  b-a  \\
b-a & 0\\
\end{pmatrix}} .
\end{equation}
The denominator could always be diagonalized and the result reads:
\begin{equation}
\frac{1}{1+i H}=\frac{1}{\begin{pmatrix}
1+i{(2\lambda)}^{1/4}\omega_{+}& 0  \\
0 & 1+i{(2\lambda)}^{1/4}\omega_{-} \\
\end{pmatrix}},
\end{equation}
where 
\begin{eqnarray}
 \omega_{+}&=&\big(c-a+\sqrt{(c-a)^2+4(b-a)^2}\,\big)/2 >0\nonumber\\
\omega_{-}&=&\big(c-a-\sqrt{(c-a)^2+4(b-a)^2}\,\big)/2<0 .
\end{eqnarray}
The analyticity domain for $\lambda$ contains at least $\cD^2$. Hence
\begin{equation}-\pi/4<  {\rm Arg} (\lambda^{1/4})<\pi/4. 
\end{equation}
It implies
 \begin{equation}
|(1+i{\lambda}^{1/4}\omega_{+})^{-1}|<\sqrt{2}\; , \;\; |(1+i{\lambda}^{1/4}\omega_{-})^{-1}|<\sqrt{2}.
\end{equation} 
So each resolvent is bounded as
\begin{equation}
\Vert \frac{1}{1+iH} \Vert \le \sqrt{2}(1+O(1/A)) .
\end{equation}
Again we shall forget the inessential factor $O(1/A)$.
Now we know that the resolvents multiply around the tree
in each contribution $Y_{\cT}$ \cite{R1}. Hence
for a tree of order $n$, the product of all the $2(n-1)$ 
resolvents in the tree is bounded by
$\sqrt2^{2(n-1)}$. The global trace adds a factor 2 to the bound so that
\begin{equation}
\left|  \Tr \prod_{{\rm around}\ \cT} \frac{1}{1+iH} \right|\le 2 \cdot \sqrt2^{2(n-1)} =  2^{n}.
\end{equation}

Now we consider the vertices.  A tree $\cT$ at $n$-th order has  $n-1$ vertices.
Each vertex contributes a factor $\sqrt\lambda$, hence we have a factor 
$\lambda^{(n-1)/2}$ in $Y_{\cT}$. There are $5$ different kinds of vertices, in the loop vertex expansion, but when considering the trace over the products of the resolvents, we only have 3 choice each time which corresponding to whether the intermediate field is $a$, $b$ or $c$. 
So the choice over the type of the vertices is bounded by an additional factor 
$3^{n-1}$. Don't forget that we have also a factor $(e^{2A})^n$ from the contour deformation. So, composing this bound with the resolvent bound we have
\begin{equation}
 \vert Y_\cT(\lambda) \vert\le2 ^{n} 3^{n-1}|\lambda|^{(n-1)/2}(e^{2A})^n\le\epsilon^{(n-1)/2} K^n 
\end{equation}
where $K=6e^{2A}$.
So we have proved this Lemma.
\end{proof}

Cayley's theorem states that the number of labeled trees over $n$ vertices is $n^{n-2}$.
Hence combining it with the Lemma we get convergence 
and analyticity of the loop vertex representation in the domain $\cD^2_{\epsilon}$:
\bea
\sum_{n=1}^{\infty}\frac{1}{n!}\ \sum_{\cT \; {\rm with }\; n\; {\rm vertices}}\ \vert Y_\cT(\lambda) \vert
&\le& \sum_{n=1}^{\infty}\frac{n^{n-2}}{n!} \epsilon^{(n-1)/2} K^n  \nonumber \\
 &\le & \sum_{n=1}^{\infty} \epsilon^{(n-1)/2} (eK)^n 
\eea
where we used Stirling's formula. This
converges for small enough $\epsilon$. 
Actually since $K=6e^{2A}$, $\epsilon = e^{-2A-2}/36$ works.

We now turn to the Taylor remainder bound. 
The remainder formula reads:
\begin{eqnarray}
&&R^N\log Z(\lambda)= 
\sum_{n=1}^{\infty}\frac{1}{n!}\ \sum_{\cT \; {\rm with }\; n\; {\rm vertices}}\ R^N Y_\cT(\lambda).
\label{sumtree1}
\end{eqnarray}
For trees with $\cT$ with $n \ge 2N+3$ we have $R^N [Y_\cT (\lambda)] = Y_{\cT}$, 
hence inserting the estimate of the previous Lemma
\bea\label{sumtree2}
\vert \sum_{n=2N+3}^\infty \frac{1}{n!}\ \sum_{\cT \; {\rm with }\; n\; {\rm vertices}} R^N Y_\cT(\lambda) \vert 
&\le& \vert \lambda\vert ^{N+1} \sum_{n=2N+3}^{\infty} \epsilon^{(n-1)/2 - (N+1)} (eK)^n \nonumber\\
&\le&\vert\lambda\vert^{N+1} K^N
\eea
for $\lambda \in \cD^2_{\epsilon}$.

So we need only to consider now trees with  $n \le 2N+2$ vertices.
Defining $\bar Y$ through $Y_\cT (\lambda) = 
\lambda^{(n-1)/2} \bar Y_\cT (\lambda)$ we have for such trees
\be
R^N Y_\cT =   \lambda^{(n-1)/2} R^{N- (n-1)/2} \bar Y_\cT, 
\ee 
and the following bound:
\begin{lemma}
In the domain $\cD^2_{\epsilon}$ we have for trees $\cT$ with $n \le 2N+2$
\bea \vert  \lambda^{(n-1)/2} R^{N- (n-1)/2} \bar Y_\cT \vert \le
\vert \lambda \vert^{N+1} K^N  \Gamma(2N - n+1) .
\eea\label{lem2}
\end{lemma}

\begin{proof}
The $R^{N- (n-1)/2}$ operator now acts on the product of resolvents
\begin{equation}
\Tr \prod_{ {\rm around}\ \cT} \frac{1}{1+iH}. 
\end{equation}
We can evaluate it through a Taylor-Lagrange
integral formula, and this formula brings intermediate fields $a$, $b$ or $c$ to the numerator.
The choice of which resolvent is derived gives a factorial but which is compensated by the factorial
in the Taylor formula itself, so these choices contribute only
$K^N$ to the bound.
Since each half vertex contributes a coupling constant $\lambda^{1/4}$, the number
of such fields brought to the numerator by the Taylor Lagrange formula must obey
\begin{equation}
(n_a+n_b+n_c)/4=N -(n-1)/2
\end{equation}
as this should be compatible with the fact that we expand to order $\lambda^{N+1}$.

Therefore we have:
\begin{eqnarray}
\vert R^{N- (n-1)/2} \bar Y_\cT \vert &\le& K^N \vert \lambda\vert^{N- (n-1)/2} \\
&&\sum_{n_a,n_b,n_c \atop n_a+n_b+n_c =4N -2n + 2}  \hskip-.7cm 
\int d\mu(a)d\mu(b)d\mu (c)
a^{n_a}b^{n_b}c^{n_c}\nonumber
\end{eqnarray}
where $d\mu(a)d\mu(b)d\mu (c)$ are the oscillating Gaussian measures 
with contour deformation for the fields $a$, $b$ and $c$ respectively. 
After using the usual bound on the resolvents and Wick contraction 
for the intermediate fields we get:
\begin{eqnarray}
\vert  \lambda^{(n-1)/2} R^{N- (n-1)/2} \bar Y_\cT \vert \le \vert \lambda\vert ^{N+1}K^N (n_a+n_b+n_c)!!
\end{eqnarray}
So the remainder is bounded in the worst case $n=0$  by:
\begin{eqnarray}
&&\vert \lambda\vert ^{N+1}K^N (n_a+n_b+n_c)!!\\
&=&\vert \lambda\vert ^{N+1}K^N(4N-2n+2)!!\le\lambda^{N+1}K(4N+2)!! \le \lambda^{N+1}K^N(2N)!\nonumber
\end{eqnarray}
where $K$ is a generic name for a constant.
\end{proof}
Combining lemmas \ref{lem1} and \ref{lem2} together with (\ref{sumtree1}) and (\ref{sumtree2})
proves that $\log Z (\lambda)$  is analytic in some $\cD^2_{\epsilon}$ domain, hence in some 
$C^2_R$ domain and that the Taylor remainder at order $N$ 
is bounded by $\vert \lambda\vert ^{N+1}K^N \Gamma(2N+1)$.
This completes the proof of Theorem \ref{theoborel3}.
\end{proof}
%%%%%%%%%%%%%%%%%%%%%%%%%%%%%%%%%%%%%%%%%%%%%%%%%%%%%%%%%%%
\section{$\phi^{2k}$ theory in zero dimension}
%%%%%%%%%%%%%%%%%%%%%%%%%%%%%%%%%%%%%%%%%%%%%%%%%%%%%%%%%%%%%%%%%%%%%%%%%%%%%%%%
\subsection{The intermediate fields for $\phi^{2k}$ theory}
In the general case of a $\lambda\phi^{2k}$ interaction, 
we could introduce the intermediate fields inductively, and in each step we attribute to the interaction term of a field $\phi$ with an intermediate field a coupling constant $\lambda^{\frac{1}{2k}}$.
In the first step we introduce a first intermediate field $\sigma_1$ and forgeting the inessential normalizing factor, and the result reads:
\begin{equation}
e^{- \lambda\phi^{2k}}=\int{d\sigma_1}e^{-\sigma_1^2/2+i\sqrt{\lambda}\phi^k\sigma_1}
\end{equation}
and
\begin{equation}  \label{gene}
 2\sqrt{\lambda}\phi^k \sigma_1=[(\lambda^{\frac{1}{2k}}\phi
\sigma_1+\lambda^{\frac{k-1}{2k}}\phi^{k-1})^2-\lambda^{\frac{1}{k}}\phi^2\sigma_1^2-\lambda^{\frac{k-1}{k}}\phi^{2k-2}  ] .
\end{equation}
For the first term in the r.h.s. we shall introduce another intermediate field $\sigma_2$ and 
we have:
\begin{equation}
 e^{i(\lambda^{\frac{1}{2k}}\phi
\sigma_1+\lambda^{\frac{k-1}{2k}}\phi^{k-1})^2}=\int d \sigma_2 e^{i(\lambda^{\frac{1}{2k}}\phi
\sigma_1+\lambda^{\frac{k-1}{2k}}\phi^{k-1})\sigma_2} e^{-i\sigma_2^2 /2}.
\end{equation}
For the second term we have simply 
\begin{equation}
e^{- i \lambda^{\frac{1}{k}}\phi^2\sigma_1^2}=\int d\sigma_3 e^{- i\lambda^{\frac{1}{2k}}\sigma_3 \phi \sigma_1} e^{i \sigma_3^2/2}.
\end{equation}
The third term has the potential $\phi^{2k-2}$ which means that we have the same type of interaction but with the degree of the potential lowered by $2$ and the coupling constant lowered by degree $\lambda^{\frac{1}{k}}$. We could repeat this process inductively until the final form is linear at most in each of the final $[3(k-2) +1]$ intermediate fields $\sigma_i$, quadratic at most in $\phi$, and trilinear in all fields taken together, which means the field $\phi $ and all the intermediate fields 
Remark that we can maintain imaginary factors throughout the induction,
by using imaginary Gaussian integrals. Again we integrate out some of the intermediate fields and the initial field $\phi$. The result could always be written in the form (up to inessential normalization constants)
\begin{equation}
Z(\lambda) = \int \prod_r da_r \prod_s db_s\prod dc\  e^{i(a_1^2- a_2^2- a_3^2+ b_1^2- b_2^2- b_3^2 \pm c^2 \cdots)/2}e^{V}
\end{equation}
where 
\begin{equation}
 V=-\frac{1}{2}\Tr\ln[A+iH(\{a\},\{b\},\{c\}...)].=-\frac{1}{2}\Tr\ln[G].
\label{fulmatrix}
\end{equation}
Here $A=diag(1,1, i, -i...)$ where the number of $1$ depends on whether $k$ is even or odd: if $k$ is even, there is only one $1$ in $A$ and the other diagonal elements are $\pm i$; 
if $k$ is odd the first two diagonal elements are $1$s and the other diagonal elements are $\pm i$. $a_i$, $b_i$, $c_i...$ represent the remaining intermediate fields.
 $H$ is a symmetric matrix with nonzero elements appearing only in the first line and the fist colum, for example:
%%%%%%%%%%%%%%%%%%%%%%%%%%%%%%%%%%%%%%%%%%%%%%%%%%%%%%%%%%%%%%%%%%%%%%
\begin{equation}
 H=\lambda^{\frac{1}{2k}}\begin{pmatrix}
\lambda^{\frac{1}{2k}}g_1( a_i, b_i, c_i...) &  g_2 ( a_i, b_i, c_i...) & g_3 ( a_i, b_i, c_i...) &... \\
g_2 ( a_i, b_i, c_i...) & 0 & 0 & ...\\
g_3( a_i, b_i, c_i...) & 0 & 0 & ...\\
...&... &... & ...\\
\end{pmatrix}.
\label{fulllinearmatrix}
\end{equation}
Here $g_j( a_i)$ is a sum of {\emph{linear}} terms in the intermediate fields that appears in the determinant.

We take the $e^{-\lambda\phi^{8}}$ model for example. In this case $k=4$, so we associate to each field $\phi$ a coupling constant $\lambda^{\frac{1}{2k}}=\lambda^{1/8}$. The interaction form can also be written as
 \begin{equation}
\int d \sigma d b_i d X  e^{-X G X^t} e^{-\frac{1}{2}\sigma^2-\frac{i}{2}(a_1^2-a_2^2-a_3^2+b_1^2-b_2^2-b_3^2)}
\end{equation}
where 
\begin{equation}
 X=\begin{pmatrix}
\phi, a_1, a_2, a_3
   \end{pmatrix}
\end{equation}
and
\begin{eqnarray}
&&G=A+iH= \begin{pmatrix}
1 & 0 & 0 & 0\\
0  & i & 0 & 0\\
0 & 0 & -i & 0\\
0 & 0 & 0 & -i\\
\end{pmatrix}  \nonumber\\&+&i\lambda^{1/8}\begin{pmatrix}
-\lambda^{1/8}(b_1-b_3) &  -(b_1+\sigma) & \sigma & b_1-b_2 \\
 -(b_1+\sigma)  & 0 & 0 & 0\\
\sigma & 0 & 0 & 0\\
b_1-b_2& 0 & 0 & 0\\
\end{pmatrix}\nonumber \\
\end{eqnarray}
where $a_i$, $b_i$ and $\sigma$ are the intermediate fields. It is not surprising that we have an element with a coupling constant $\lambda^{1/4}$ in the matrix, as this term corresponds to the interaction term $\phi^2 (b_1-b_3)$, and we associate to each $\phi$ a factor $\lambda^{1/8}$: We have a similar 
situation for all other $\phi^{2k}$ case, see (\ref{fulllinearmatrix}).

Then we consider a more complicated example, the $\exp(-\lambda\phi^{10})$ model. In this case we have $k=5$ and the coupling constant for each field $\phi$ is $\lambda^{1/10}$. The interaction form can be written as
 \begin{equation}
\int  d a_i d c_i d X  e^{-X G X^t} e^{-\frac{i}{2}(a_1^2-a_2^2-a_3^2+c_1^2-c_2^2-c_3^2)},
\end{equation}
 where 
\begin{equation}
X=\begin{pmatrix}
\phi, \sigma, b_1, b_2, b_3
   \end{pmatrix}
\end{equation}
and
\begin{eqnarray}
&&G=A+iH= \begin{pmatrix}
1 & 0 & 0 & 0 & 0\\
0 & 1 & 0 & 0 & 0\\
0 & 0 & i & 0 & 0\\
0 & 0 & 0 & -i & 0\\
0 & 0 & 0 & 0 & -i\\
\end{pmatrix}\\
&+&i\lambda^{1/10}\begin{pmatrix}
-\lambda^{1/10}(c_1-c_3) &  -(a_1-a_2) & -(a_1+\sigma) & a_1-a_3 & c_1-c_2 \\
  -(a_1-a_2) & 0 & 0 & 0 &0\\
 -(a_1+\sigma)& 0 & 0 & 0 & 0\\
 a_1-a_3& 0 & 0 & 0 & 0\\
c_1-c_2& 0 & 0 & 0 & 0\\
\end{pmatrix}.\nonumber
\end{eqnarray}
where $a_i$, $b_i$,  $c_i$ and $\sigma$ are the intermediate fields. 

\begin{lemma}
The inverse of the matrix $G$ exists and is bounded by $\sqrt{2}$.
\end{lemma}

\begin{proof}
The matrix $G=A+iH$ is a symmetric matrix that has only non 
zero elements in the first line, the first row and the
diagonal. We have
\begin{equation}
 G=A+iH=A(\bbbone+iA^{-1}H) .
\end{equation}
As $A$ is a diagonal matrix whose elements are either $1$ or $\pm i$, the inverse of $A$ is bounded and has a similar structure. So in the following we consider only the inverse of the matrix $\bbbone+iA^{-1}H$.
For a general $\phi^{2k}$ theory we have
\begin{equation}
iA^{-1}H= i \lambda^{\frac{1}{2k}}\begin{pmatrix}
\lambda^{\frac{1}{2k}}d_1 & d_2 & d_3 &...&...&d_n\\
\pm i d_2 & 0 & 0 & ...&...&0\\
\pm i d_3 & 0 & 0 & 0 & ...& 0\\
... & ... &  ..&.... & ...&...\\
\pm i d_n & 0 & 0 & 0 &0 & 0\\
\end{pmatrix}.
\end{equation}
where $d_i$ is an arbitrary element of $A^{-1}H$ which has non vanishing elements only in the first row and first column. The matrix $A^{-1}H$ has only two non vanishing eigenvalues, each of multiplicity 1. The exact formula for the eigenvalues depends also on whether $k$ is even or not. For $k$ even, we have
\begin{eqnarray}
 \omega_{\pm}=\frac{-\lambda^{\frac{1}{2k}} d_1(1\pm \sqrt{1-\frac{4i\lambda^{-1/k}B}{d_1^2}})}{2}
\end{eqnarray}
where 
\begin{equation}
B=\pm d_2^2 \pm d_3^2 \pm...\pm d_n^2
\end{equation}
is a combination of the squares of the intermediate fields.

When $k$ is odd, we have
\begin{equation}
 \omega_{\pm}=\frac{-\lambda^{\frac{1}{2k}} d_1(1\pm \sqrt{1+\frac{4\lambda^{-1/k}}{d_1^2}(d_2^2-iB')})}{2}
\end{equation}
and in this case 
\begin{equation}
B'= \pm d_3^2 \pm...\pm d_n^2 .
\end{equation}

Through some basic calculation we can easily find that in both case we have 
\begin{equation}
| 1+i\omega_{\pm}|\geq \frac{1}{\sqrt{2}} .
\end{equation}
So we have
\begin{eqnarray}
 \bbbone+iA^{-1}H=  P \begin{pmatrix}
1+i\omega_{+} & 0 & 0 & 0 &...& 0\\
0 & 1+i\omega_{-}  & 0 & 0 & ...&0\\
0 & 0 & 1 & 0 &...& 0\\
...& ... & ...& ... & ...&...\\
0 & 0 & 0 & 0 &...& 1\\
\end{pmatrix} P^{-1}  \\
\end{eqnarray}
where $P$ is the diagonalizing matrix. Therefore $ \bbbone+iA^{-1}H$
is invertible and its inverse has eigenvalues $(1+i\omega_{\pm})^{-1}$ and
1 so that we have
\begin{equation}
 \parallel [\bbbone+iA^{-1}H]^{-1} \parallel \leqslant
\sqrt{2} .
\end{equation}

This proves this lemma.
\end{proof}

%%%%%%%%%%%%%%%%%%%%%%%%%%%%%%%%%%%%%%%%%%%%%%%%%%%%%%%%%%%%%
\subsection{The analytic domain and contour deformation}
In the $\lambda\phi^{2k}$ theory, the analytic domain for the coupling constant $\lambda$ is
\begin{equation}
 -\frac{(k-1)\pi}{2}\leqslant {\rm Arg\; }  \lambda\leqslant-\frac{(k-1)\pi}{2}.
\end{equation}

As for each $\phi$ we have a coupling constant $\lambda^{\frac{1}{2k}}$, and as in the matrix $G$ each element is linear in $\phi$,  we have for each element $a_i$ in $G$ the relation:
\begin{equation}
 -\frac{(k-1)\pi}{4k} \leqslant {\rm Arg\; } {a_i}\leqslant \frac{(k-1)\pi}{4k}.
\end{equation}
 Similarly we could prove that the inverse of the matrix $G$ is bounded for all $\lambda$ in its analytic domain. To be more precisely, we have 
\begin{equation}
\vert 1+i\omega_{\pm} \vert \geq c\sin {\frac{\pi}{2k}} 
\end{equation}
for either $k$ even or odd, with $c$ a small constant. So we proved 
that the inverse of matrix $G$ exists and is bounded.

The contour deformations then proceeds as in the previous section, since all
the intermediate fields integrals are of the same type and we get again a
bound of the type (\ref{contourbound}).

\subsection{Borel summability }
The proof of the Borel summability  for $\phi^{2k}$ theory is quite similar to the $\phi^6$ theory. We shall first of all consider the Borel summability  for the partition function $Z$ and then the connected function $\log Z$.
\begin{theorem}
The partition function of a field theory 
with potential $\lambda\phi^{2k}$ is Borel summable or order $k-1$.
\end{theorem}
\begin{proof}
This theorem is easy and do not need any loop vertex expansion. In this case the analytic region for $\lambda$ would become $\cD^{k-1}=\{-\frac{(k-1)\pi}{2}<Arg(\lambda)<\frac{(k-1)\pi}{2}\}$. 
The argument for analyticity and Taylor bounds is the same as above, replacing $2$ by $(k-1)$.
\end{proof}

\begin{theorem}
 The connected function $\log Z(\lambda)$ for the theory with potential $\lambda \phi^{2k}$ is  Borel-Le Roy summability at order $k-1$.
\end{theorem}
\begin{proof}
The argument is quite similar to the $\lambda\phi^6$ case and we need the loop vertex expansion.
In (\ref{gene}) we have shown that the general potential $\lambda\phi^{2k}$ could always be expressed in terms of intermediate fields. After each step we have a new potential $i\phi^{2k-2}$. We could get the bound for the connected function with the same method as in the $\phi^6$ case.
Now we consider the factorials.
In the intermediate fields expression, each intermediate field is linearly interacting with a field $\phi$ and coupling constant $\lambda^{\frac{1}{2k}}$, so after the expansion to $N$-th order of the coupling constant $\lambda$, and Wick contraction, we get a factor
\begin{equation}
\frac{[2kN]!!}{N!}\sim (k-1)N!.
\end{equation}
Combining all the arguments above we find that the remainder of the Taylor expansion is bounded by 
\begin{equation}
|\lambda|^{N+1}K^N\Gamma[(k-1)N+1]
\end{equation}

So the connected function is Borel -Le Roy summable of order $k-1$.
\end{proof}

\section{Conclusion and Perspectives}

It is now clear that the traditional constructive tool of decomposing 
space into an \emph{ad hoc} lattice of cubes and performing cluster expansions
is not fundamental and can be replaced by better techniques.
The loop vertex expansion \cite{R1} and \cite{MR1} is the first of these. A 
different but related approach is proposed in \cite{GMR}. 
The fundamental idea of the loop vertex expansion is to decompose an interaction of arbitrary degree until
trilinear or ``three body'' interactions are reached, since these are the most ``basic''. 
The basic objects are loops made out of a subfamily of the corresponding fields.
The loops are made of resolvents, which are uniformly bounded in the case of stable interactions, and
they are joined by explicit propagators into cacti structures.
This technique reconciles constructive theory and the spirit of Feynman's perturbative theory. The
essential mathematical problem of field theory is to iteratively compute connected functions in order to perform renormalization.
In Feynman's graphical representation of field theory, connected functions 
were very easy to compute since they were written as sums of connected graphs, but the corresponding
formulas have no mathematical meaning since the expansion diverges. In the loop vertex expansion formalism connected functions are still very easy to identify as they are written as sums of connected cacti, but these sums are now convergent, hence the corresponding formulas are mathematically meaningful.

It will become increasingly necessary in our opinion to develop advanced 
constructive techniques such as loop vertex expansions to understand nonperturbatively
new field theories such as non commutative field theories or group field theories of
quantum gravity. Indeed these theories include non-local aspects which, up to our knowledge,
cannot be treated through lattice of cubes decomposition and traditional cluster expansions.

Still a long road is to be performed to validate this new constructive philosophy and to push it
up to the level where we can reproduce with it all the previous results of the constructive
literature over the last decades. This paper accomplished a small but significant step
in showing that the decomposition into trilinear interactions does not work
solely for the $\phi^4$ interaction but also for interactions of any degree.
But clearly the limitation to zero dimension must now be lifted. The three main steps ahead
are the construction of models in single renormalization group slice, the construction
of matrix models with arbitrarily high degree interaction and correct scaling
as the size of the matrix gets large, and finally the inclusion of renormalization.

%%%%%%%%%%%%%%%%%%%%%%%%%%%%%%%%%%%%%%%%%%%
\subsection{Sliced $\phi^{2k}$model in any dimension}
We could easily generalize the loop vertex expansion method to a $\phi^{2k}$ theory in any dimension
in a single renormalization group slice, by following \cite{MR1}. We only sketch the general idea
in this paper, details being devoted to a future publication. For instance
In $D$ dimensions the propagator in a single renormalization group slice reads:
\begin{equation}
C_j(x, y)=\int_{M^{-2j}}^{M^{-2j+2}}e^{-\alpha m^2}e^{-(x-y)^2/{4\alpha}}\alpha^{-D/2} d\alpha\leqslant KM^{(D-2/2)j}e^{-cM^j|x-y|}
\end{equation}
In $\phi^{2k}$ we associate to each field $\phi$ a coupling constant $\phi^{1/2k}$ and an operator $D_j=C_j^{1/2}$.  And we still have 
\begin{equation}
 \int d_{\mu_{C_j}}(\phi)e^{-\int\lambda\phi^{2k}}=\int d \nu(\sigma_i)e^{-\frac{1}{2}log(A+iH)}
\end{equation}
where $A$ is the same as in the formula (\ref{fulmatrix}), and
\begin{equation}
  H=\lambda^{\frac{1}{2k}}\begin{pmatrix}
\lambda^{\frac{1}{2k}}  D_j g_1( a_i, b_i, c_i...)D_j &  g_2 ( a_i, b_i, c_i...)D_j & g_3 ( a_i, b_i, c_i...)D_j &... \\
D_jg_2 ( a_i, b_i, c_i...) & 0 & 0 & ...\\
D_jg_3( a_i, b_i, c_i...) & 0 & 0 & ...\\
...&... &... & ...\\
\end{pmatrix}.
\label{ful2d}
\end{equation}
We find that the form of $H$ is quite similar to the formula (\ref{fulllinearmatrix}) except that to each factor $\lambda^{\frac{1}{2k}}$ is now associated a factor $D_j$ and  we require that the first column is the transpose conjugate to the first row, as $D_j$ are all operators now.

Then the proof of the uniform Borel summability and the decay of connected functions
should follow in the same way as in \cite{MR1}. 

\subsection{Matrix models}

A very interesting property of loop vertex expansions is to allow
uniform Borel summability theorems on matrix
models with the right scaling of the Borel radius as the matrix gets large \cite{R1}.

To extend this result to a single matrix model with $\phi^{2k}$ interaction and matrix of size $N$
we should prove Borel-Le Roy summability with a radius that scales like $N^{-(k-1)}$. This seems
doable but all the intermediate fields are now matrix like and
one should carefully control the normalization factors associated to contour deformation 
of the corresponding fields in section \ref{contourdefo}.

\subsection{Renormalization}
This is the most difficult part. The first goal should be eg to construct very simple models
such as the $\phi^4_2$ model which requires only Wick ordering with the loop vertex 
expansion technique. Then we expect the loop vertex expansion should be applied to
just renormalizable models such as infrared $\phi^4_4$ \cite{GK,FMRS} and ultimately it should be a key tool for the hopefully full construction of an interacting field theory in four dimensions, namely
the Grosse-Wulkenhaar model \cite{GW1,RVW,DGMR,GW2}.

%%%%%%%%%%%%%%%%%%%%%%%%%%%%%%%%%%%%%5%%%%
\medskip
\noindent{\bf Acknowledgments}
We thank Jacques Magnen and Alan Sokal for useful discussions or suggestions.

\end{document}